\documentclass[twocolumns,10pt]{emulateapj}
\usepackage{amsmath, amssymb}
\usepackage{graphicx}
\usepackage{latexsym}
\usepackage{dcolumn}
\usepackage{color}

\newcommand{\beq}{\begin{equation}}
\newcommand{\beqa}{\begin{eqnarray}}
\newcommand{\eeq}{\end{equation}}
\newcommand{\eeqa}{\end{eqnarray}}
\def \bmath #1 {{\hbox{\boldmath{$#1$}\unboldmath}}}

\shorttitle{Formation and destruction of SiS in space}
\shortauthors{Zanchet  et al.}

\begin{document}
 
\title{Formation and destruction of SiS in space}

\author{ Alexandre Zanchet$^{1,\dag}$, Octavio Roncero$^1$,
  Marcelino Ag\'undez$^1$, and Jos\'e Cernicharo$^1$ }                                                
\email{octavio.roncero@csic.es}
\affil{
$^1$ Instituto de F{\'\i}sica Fundamental, CSIC, C/ Serrano 123, E-28006 Madrid, Spain\\
$^\dag$ Present address:
    Facultad de Ciencias Qu\'imicas, Universidad Complutense de Madrid, Plaza de las Ciencias,
    Ciudad Universitaria, E-28040 Madrid, Spain
    }
\begin{abstract}

  The presence of SiS in space seems to be restricted to a few selected types of astronomical environments.
  It is long known to be present in circumstellar envelopes around evolved stars and it has also been detected
  in a handful of star-forming regions with evidence of outflows,
  like Sgr\,B2, Orion\,KL and more recently L1157-B1.
  The kinetics of reactions involving SiS is very poorly known and here we revisit
  the chemistry of SiS in space by studying some potentially important reactions of formation
  and destruction of this molecule. We calculated {\it ab initio} potential energy surfaces
  of the SiOS system and computed rate coefficients in the temperature range 50-2500 K for the reaction of destruction of SiS,
  in collisions with atomic O, and of its formation, through the reaction between Si and SO.
  We find that both reactions are rapid, with rate coefficients of a few times
  10$^{-10}$ cm$^3$ s$^{-1}$, almost independent of temperature.
  In the reaction between Si and SO, SiO production is 5-7 times more efficient than SiS formation.
  The reaction of SiS with O atoms can play an important role in destroying SiS in envelopes
  around evolved stars. We built a simple chemical model of a postshock gas to study the
  chemistry of SiS in protostellar outflows and we found that SiS forms with a lower abundance
  and later than SiO, that SiS is efficiently destroyed through reaction with O,
  and that the main SiS-forming reactions are Si + SO and Si + SO$_2$.

  \vspace{1cm} Accepted in Astrophysical journal , ApJ (2018)
\end{abstract}



\section{Introduction} \label{sec:introduction}

The molecule SiS was first observed in the molecular envelope
around the carbon-rich AGB star IRC\,+10216 and in the molecular cloud
toward the Galactic Center Sgr\,B2 \citep{Morris1975}. This species is
known to be present in circumstellar envelopes around evolved stars,
where it is thought to be formed under thermochemical equilibrium
conditions at the stellar surface and later on ejected into the expanding
wind \citep{Bujarrabal1994,Cernicharo2000,Schoier2007,Agundez2012,Fonfria2015,Velilla-Prieto2015}.
The molecule is observed in both carbon-rich and oxygen-rich envelopes,
but SiS maser emission is only observed in the carbon-rich object IRC\,+10216 \citep{Fonfria2006}.
Apart from the envelopes of evolved stars, SiS has only been found in a handful of star-forming regions
with evidence of outflows, like Sgr\,B2 and Orion\,KL \citep{Dickinson1981,Ziurys1988,Ziurys1991,Tercero2011}.
Recently, \citet{Podio2017} reported the detection of SiS in L1157-B1,
a shocked region associated with an outflow driven by a low-mass protostar.

Despite SiS is a stable molecule involving two elements with high cosmic abundances,
it is much less commonly observed than its oxygen chemical analogue SiO.
While SiO is prevalent in almost every outflow affected by shocks,
and as such it is considered a good tracer of shock activity \citep{Martin-Pintado1992},
the number of SiS detections in shocked outflows is scarce.
There are various obvious questions regarding the different prevalence of SiO and SiS
in the interstellar medium, in general, and in shocked outflows, in particular.
Is SiS an elusive molecule which is only formed under very particular conditions?
Could it be that lines of SiS have not been targeted as often as those of SiO,
meaning that there has been an observational bias and that both SiO and SiS are ubiquitous in shocked gas,
SiS being merely less abundant than SiO?
Answers to these questions may come from more sensitive searches for SiS
in regions with intense SiO emission. Also chemical models may hold the clue for the apparent
different behavior of SiS and SiO. However, the kinetics of chemical reactions involving SiS
is very poorly known, which hampers a good understanding of the chemistry of this molecule in interstellar space.

In this article we study various potentially important processes that can drive
the formation and destruction of SiS in interstellar and circumstellar media.
In a recent work, \cite{Rosi-etal:18} proposed the formation of this molecule
through the reactions SiH + S and SiH + S$_2$ and carried out electronic structure
calculations of the stationary points along the reactive potential energy surfaces.
In this work we study a different chemical route, the formation of SiS via the reaction
Si + SO and its destruction through the SiS + O reaction. For this purpose,
we calculate the {\it ab initio} potential energy surfaces (PESs) of the SiOS
system in the two lower electronic states and perform Quassi-classical trajectory (QCT)
calculations of the rate coefficients for several vibrational states of the reactants
in the 50-2500 K temperature range. These reaction rate coefficients are then used
in a chemical model to evaluate their impact on the formation and destruction of SiS in outflows.
The manuscript is organized as follows. First, in Sec.~\ref{sec:molecular_calculations}
we describe the calculations carried out. In Sec.~\ref{sec:astrophysics}
we discuss the impact of the studied processes on the chemistry of SiS in outflows.
Finally, in Sec.~\ref{sec:conclusions} we present our conclusions.

\vspace{0.5cm}
\section{Molecular Calculations} \label{sec:molecular_calculations}

\subsection{SiOS potential energy surfaces}

\begin{figure}
\centering
\includegraphics[angle=0,width=0.9\columnwidth]{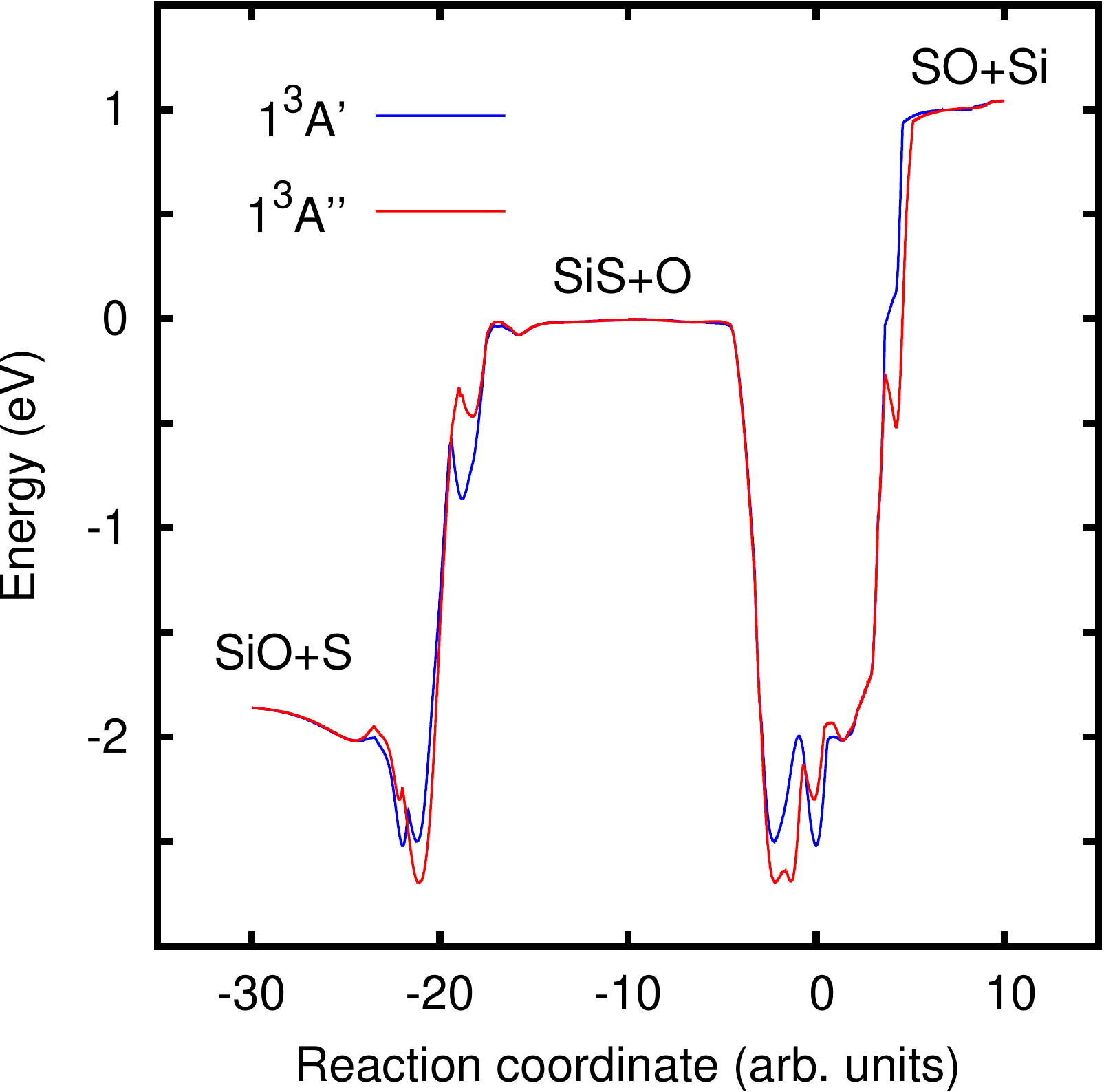}
\caption{Minimum energy paths of the two lower electronic
  states of the SiOS system,
  connecting the S($^3P$) + SiO($^1\Sigma^+$),
  SiS($^1\Sigma^+$) + O($^3P$), and SO($^3\Sigma^-$)
  + Si($^3P$) rearrangement channels.
{ The reaction coordinate is defined for the two reactions paths as
  as $r_{SiS}-r_{SO}$ and as $r_{SiO}-r_{SiS}-20$ (distances 
  in bohr). The potential energies have being optimized
  as a function of the remaining 2 coordinates in each reaction path.
  It should be noted the SO($^3\Sigma^-$) and  SiO($^1\Sigma^+$) are directly
  connected by the potential wells of the two electronic states.}
}\label{meps} 
\end{figure}

The SiOS system correlates with the open shell fragments SO($^3\Sigma^-$) + Si($^3P$),
SiS($^1\Sigma^+$) + O($^3P$), and SiO($^1\Sigma^+$) + S($^3P$), as
shown in Fig.~\ref{meps}.
All the atomic fragments are open shell $^3P$ states,
and there are several electronic states correlating to them.
In the present study we neglect spin-orbit couplings,
and only three states in each asymptote will be considered here
(correlating to the $P$ states of Si, S and O atoms, respectively).
These 3 states are classified according to their symmetry by reflection through
the plane of the molecule: two of $^3A''$ symmetry and one of $^3A'$ symmetry.
The excited 2$^3A''$ state presents a very  high barrier for the reaction,
and will not be considered in this work. The minimum energy paths connecting
the three rearrangement channels for the 1$^3A''$ and 1$^3A'$ states are shown
in  Fig.~\ref{meps}, and are clearly connected by rather deep wells.
These paths are obtained from the full-dimension PESs described below.

The PESs of the 1$^3A''$ and 1$^3A'$ states have been calculated
with a internally contracted multireference configuration interaction method
with simple and double excitations (icMRCI) \citep{Werner-etal:88},
including the Davidson correction \citep{Davidson:75}.
Aug-cc-pVTZ basis set have been considered for all atoms,
including {\it spdf} basis functions,
and the calculations have been done with the MOLPRO suite of programs \citep{MOLPRO_brief}.
In order to get an accurate and homogeneous description of the PESs,
the molecular orbitals and reference configurations were determined with a
state-average complete active space (SA-CASSCF) method \citep{Werner-Knowles:85}
for the calculation of  the first two  $^3A'$ and the first two $^3A''$ electronic states.
An active space of 12 electrons in 10 orbitals (12-18{\it a'} and 3-5{\it a''})
were used in order to include all the relevant valence orbitals with affordable calculation time.
The 1s orbital of oxygen were kept frozen as well as the 1s, 2s and 2p orbitals of Si and S.

{\it Ab initio} icMRCI+Q calculations were performed for 3042 geometries for the 1$^3A'$ and 1$^3A''$ states.
These points have been fitted using a method based on the Reproducing Kernel Hilbert Space (RKHS)
for each adiabatic state independently. This procedure is known to be well adapted
for potentials of triatomic systems with complex topologies consisting of several wells
and saddle-points \citep{Zanchet-etal:06,Zanchet-etal:09}.
In addition, a fast evaluation algorithm can be implemented when points are sampled
on a regular grid \citep{1997-TH-JCP-7223} like in this work,
making this kind of analytical potentials very appropriate to perform dynamics calculation.

\begin{figure*}
\centering
\includegraphics[angle=0,width=0.9\columnwidth]{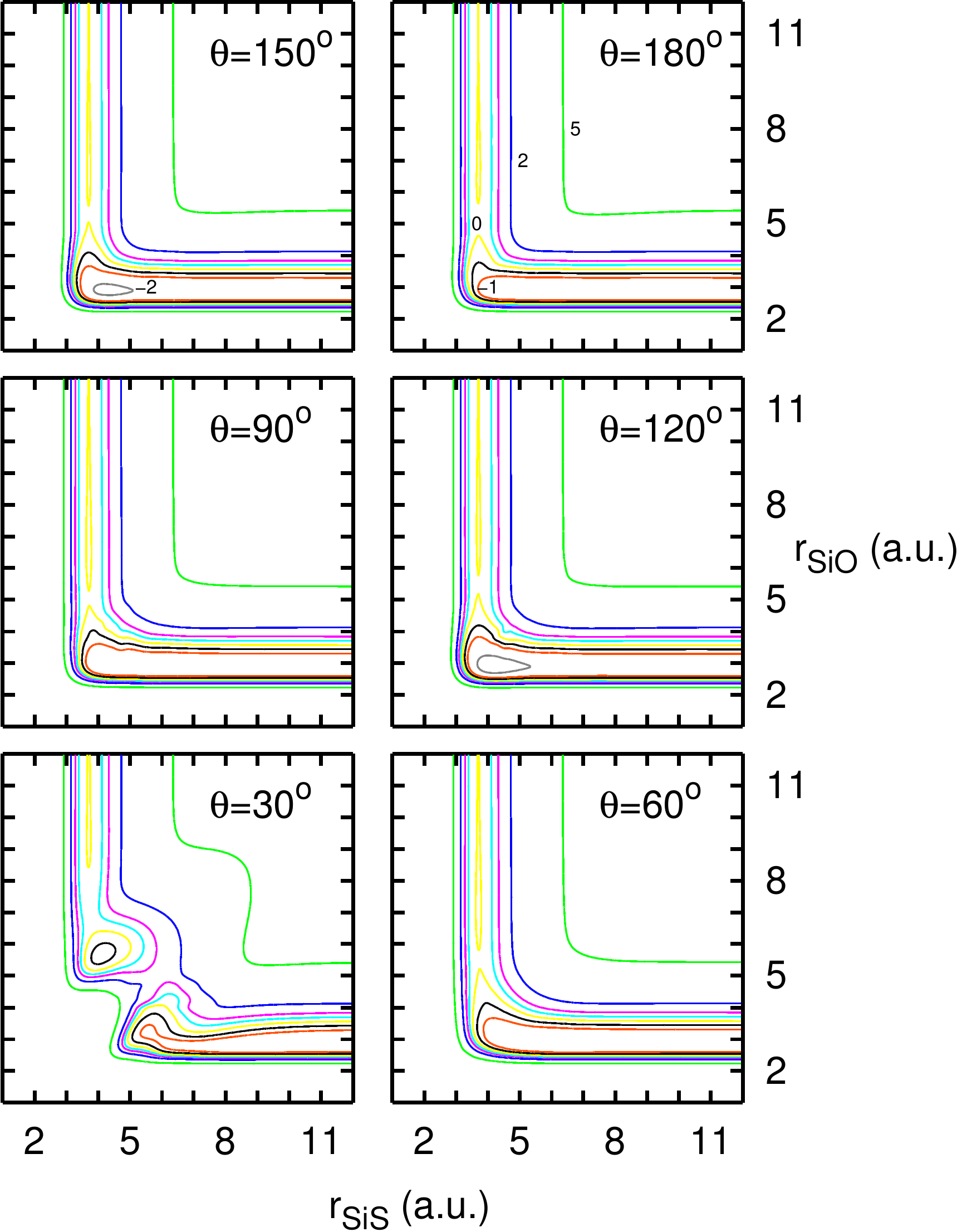} \hspace{1.5cm} \includegraphics[angle=0,width=0.9\columnwidth]{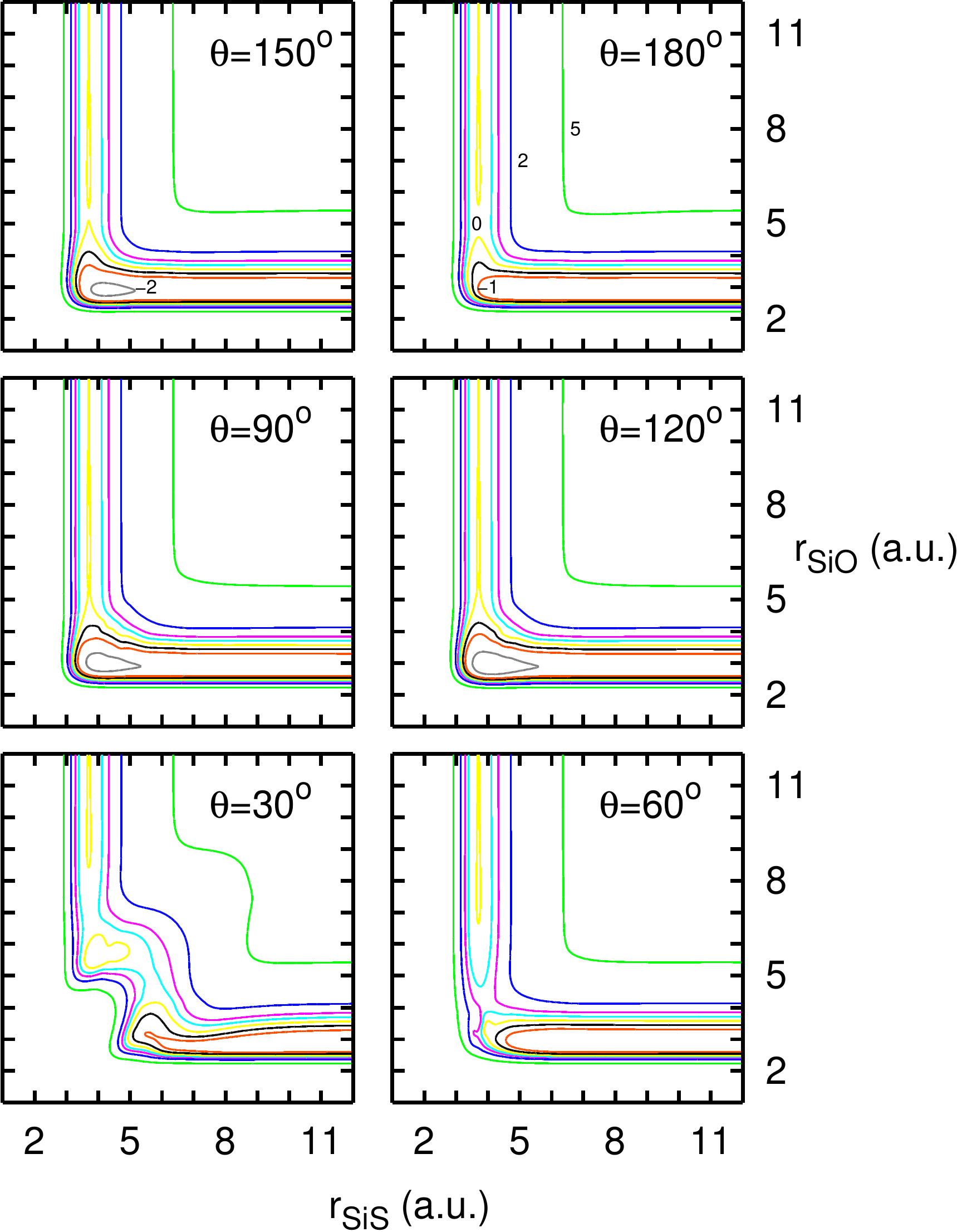}
\caption{Contour plots of the potential energy surfaces of SiOS($^3A''$), left panels,
  and SiOS($^3A'$), right panels, as a function of the SiS and SiO distances in atomic units.
  Each panel corresponds to an angle $\theta$ between the SiS and SiO vectors, as indicated in each panel.
  The energies of the countours are 5, 2, 1, 0.5, 0, -0.5, -1 and -2 eV.} \label{SiSOpes}
\end{figure*}

The two PESs are shown in Fig.~\ref{SiSOpes} and are very similar:
they present 2  deep wells associated to SSiO and OSSi isomers,
and a shallow van der Waals well in the entrance channel.
The saddle-point between the van der Waals and the SSiO wells give rise to a submerged barrier
in the entrance channel (see Fig.~\ref{SiSOpes}).
At these submerged barriers, the variation of the potential with the SiS stretching distance
becomes narrower than for free SiO($^1\Sigma^+$).
This will have interesting effects in the dynamics described below.
A more detailed description of the fits and the PESs will be given elsewhere (Zanchet et al., in preparation).

In this work we focus on two reactions. The formation of SiS($^1\Sigma^+$) from Si($^3P$) + SO($^3\Sigma^-$),
which competes with the more exothermic channel leading to SiO($^1\Sigma^+$)  + S($^3P$),
\begin{eqnarray} \label{formation-reaction}
Si(^3P)+SO(^3\Sigma^-) &\rightarrow& SiS(^1\Sigma^+) + O(^3P) + 1~{\rm eV} \\
                                        &\rightarrow& SiO(^1\Sigma^+) + S(^3P) + 2.9~{\rm eV}, \nonumber
\end{eqnarray}
and the destruction of SiS($^1\Sigma^+$) by the reaction
\begin{eqnarray} \label{destruction-reaction}
SiS(^1\Sigma^+) + O(^3P) \rightarrow SiO(^1\Sigma^+) + S(^3P) + 1.9~{\rm eV}.
\end{eqnarray}
These reactions are exothermic and are discussed below.
{ They correspond to the first triplet states on the three arrangement channels.
    It is worth mentioning that  the Si($^3P$)+SO($^3\Sigma^-$) correlates not only
    with these triplet states, but also with one singlet  and five quintuplet states.
    The quintuplets correlate with very high states in the SiS and SiO asymptotes and their
    contribution to the reactions studied here can be neglected. The singlet correlates
    with the other two arrangment channels as
    \begin{eqnarray} \label{formation-reaction-singlets}
            Si(^3P)+SO(^3\Sigma^-) &\rightarrow& SiS(^1\Sigma^+) + O(^1D) - 1~{\rm eV} \\
                                        &\rightarrow& SiO(^1\Sigma^+) + S(^1D) + 1.75~{\rm eV}. \nonumber
    \end{eqnarray}
    SiS($^1\Sigma^+$) + O($^1D$) is endothermic by $\approx$ 1eV, and it is closed for the
    energies considered. The  SiO($^1\Sigma^+$) + S($^1D$) channel  is open, and can be formed,
    but since the statistical weight of the singlet is one third with respect to that of
    the triplets, its contribution is expected to be small, and will not be considered in the present work.
}

\subsection{SiS destruction: SiS + O $\rightarrow$ SiO+ S} \label{subsec:sis+o}

\begin{figure}[b]
\centering
\includegraphics[angle=0,width=\columnwidth]{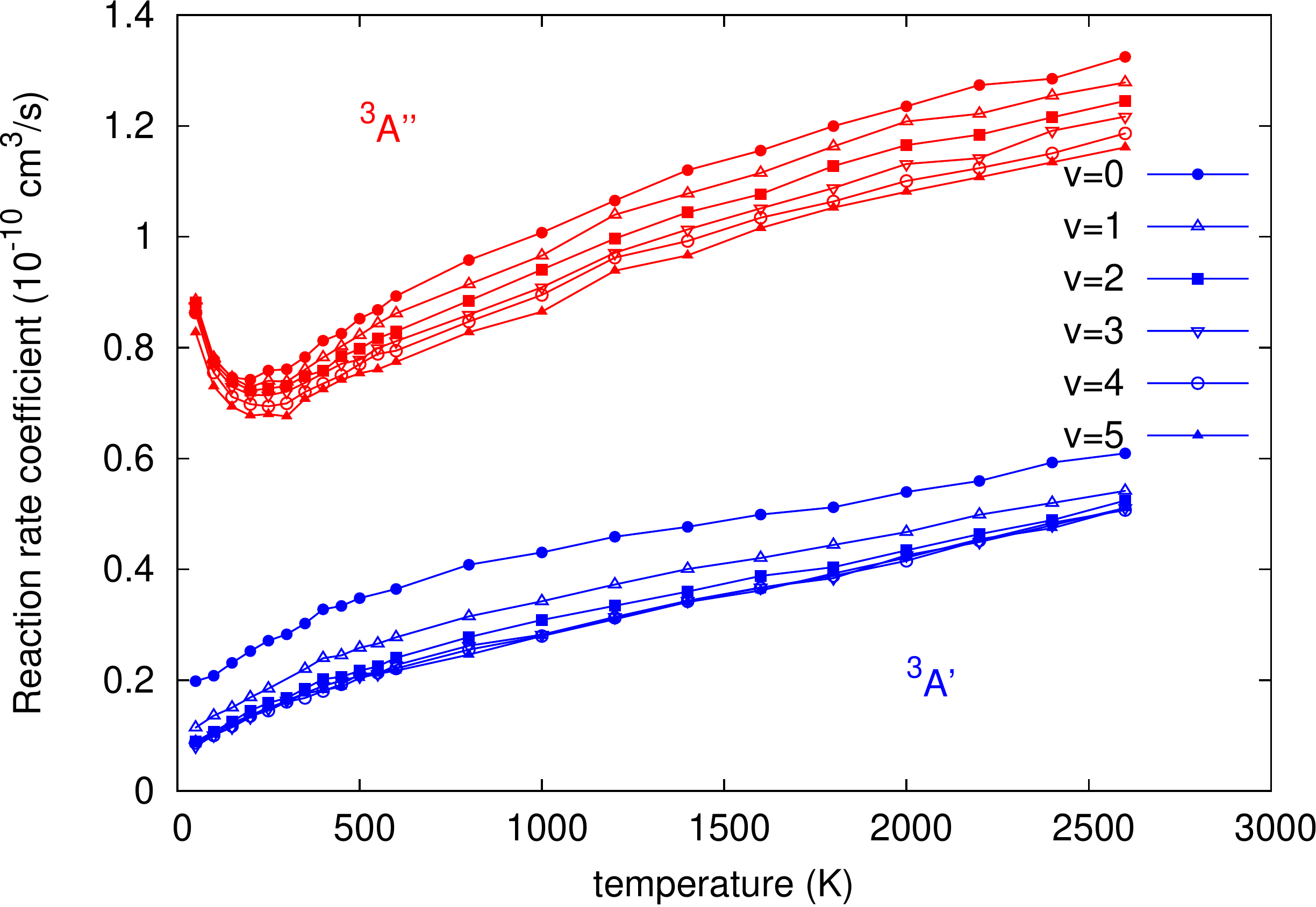}
\caption{Vibrational state selected rate coefficients, $K_v$,
  obtained in the QCT calculations of the SiS($^1\Sigma^+, v$) + O($^3P$)
  reactive collisions described in the text for the two 1$^3A''$ and 1$^3A''$
  electronic states and different vibrational levels as indicated in the caption by different symbols.} \label{SiSOrates}
\end{figure}

The use of quantum methods in systems with such heavy atoms and deep wells is still a challenge nowadays.
However, classical mechanics is expected to give rather accurate results,
and for this reason in this work we use the QCT method \citep{Karplus-etal:65}
as implemented in the miQCT code~\citep{Dorta-Urra-etal:15,Zanchet-etal:16}.
For each vibrational state of SiS, $5\times10^5$ trajectories are run changing the initial conditions,
consistent with a Boltzmann distribution of translation and rotation energy at a given temperature $T$.
The initial distance between reactants is set to 50 a.u., with a maximum impact parameter of 35 a.u.
The trajectories are stopped when any distance becomes larger than 55 a.u.
A conservation of energy better than 0.01 meV is imposed.
From these calculations, the vibrational state selected rate coefficient,
$K_v$, are calculated and are shown in Fig.~\ref{SiSOrates}.{ The rate for each electronic states, $^3A'$ and $^3A''$,
  has been multiplied by 1/3, to account for the orbital angular momentum
  degeneracy, while the spin degeneracy is absent because all triplet
  states lead to the same reactivity}

\begin{figure}[b]
\centering
\includegraphics[angle=0,width=\columnwidth]{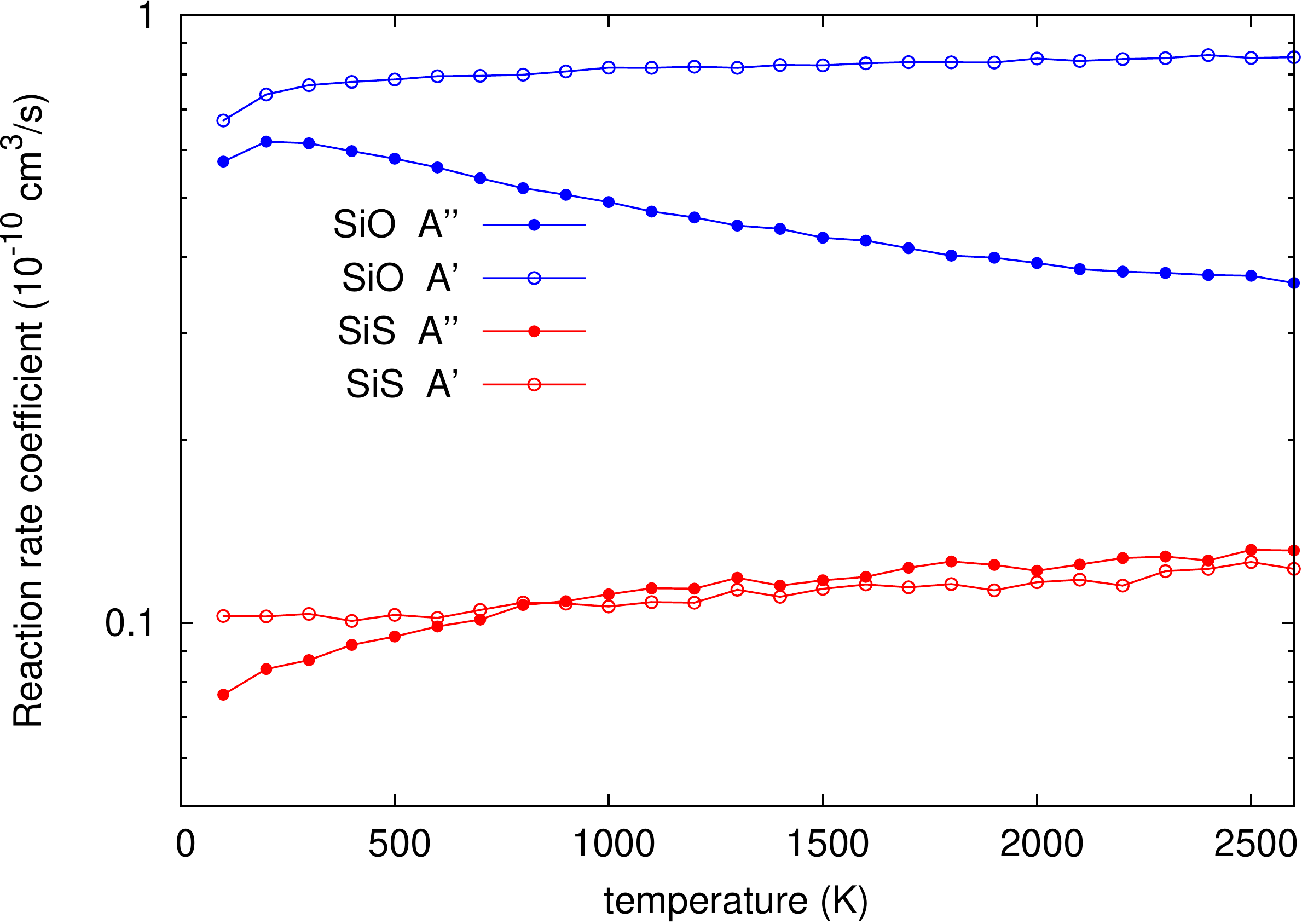}
\caption{Vibrational state selected rate coefficients,
  $K_v$, obtained in the QCT calculations for the SO($^3\Sigma^-$,v=0) + Si($^3P$)
  collisions to form SiS + O($^3P$) and  SiO + S($^3P$) products for
  the two 1$^3A''$ and 1$^3A''$ electronic states.} \label{SO+Sirates}
\end{figure}

The rate coefficients for the $1^3A''$ state are 3-4 times larger than those obtained for the $1^3A'$.
This is due to the larger cone of acceptance ($i.e.$ the amplitude of
the configuration space between reactants and products) of the 1$^3A''$ state simply
because the submerged barrier in the entrance channel is lower in energy.
In all cases, the rate coefficient increases with increasing translational temperature,
also as a consequence of the increase of cone of acceptance. It is interesting to note that 
the rates coefficient decreases with increasing the initial vibrational excitation of SiS for the two electronic state.
This behavior is due to the narrowing of the PES 
as a function of SiS internuclear distance in the entrance channel.
As can be seen in Fig.~\ref{SiSOpes}, it occurs for SiO distances of $\approx$ 5 a.u.,
prior to reach the van der Waals well.
Finally, in the case of the 1$^3A''$ state the rate coefficients increase for $T< 300 $K,
which is interpreted as a manifestation of a complex forming mechanism occurring
at these low temperatures due to the van der Waals well in the reactants channel.
A similar rise is expected for the 1$^1A'$ state but at lower temperatures.
A more careful description of the reaction dynamics is now in progress (Zanchet et al., in preparation).

\subsection{SiS formation: Si + SO $\rightarrow$ SiS + O}

The formation of SiS($^1\Sigma^+$) has also been analysed by performing
a QCT study of the reactive collision between Si and SO.
The method and analysis are similar to those discussed in Sec.~\ref{subsec:sis+o}.
The difference is that in this case there are two different products: SiS and SiO.
The rate coefficients calculated for each of these two products are shown
in Fig.~\ref{SO+Sirates}.
{ This asymptote correlates with 1 singlet, 3 triplets and 5 quintuplets,
  and  the three $P$ states of Si, making a total of 27 states.
  Neglecting the contributions
  of singlet and quintuplets, and considering the triple degeneracy of triplets,
  the total electronic partition function is
  3/27 for $^3A'$ and $^3A''$,  included in Fig.~\ref{SO+Sirates}.
}  
The formation rate of SiO is always larger than that of SiS.
The SiO/SiS ratio between these two rate coefficients changes with
  temperature being about 7 at 500 K and $\approx$5 at 2000 K.
This reduction of the ratio with increasing temperature can be explained by statistical arguments:
the number of accessible states for SiS increases faster than for SiO because SiS
vibrational frequencies and rotational constants are lower than those of SiO.

The total reaction rate coefficients obtained for the formation and destruction rates (summed
over electronic and vibrational states) have been fitted to the usual expresion
$$
K(T)= \alpha\,\left({T\over 300}\right)^\beta\, \exp(-\gamma/T),
$$
and the $\alpha$, $\beta$ and $\gamma$ parameters are listed in Table I,
together with those corresaponding to other reactions used in the astrochemical model
described below.

\section{Astrochemical impact} \label{sec:astrophysics}

The kinetics of chemical reactions involving SiS is very poorly known and the processes
studied here are potentially important in the formation and destruction of SiS in space.
Because we have quantitatively evaluated the kinetics of the studied processes,
we can draw some conclusions about their role in the chemistry of SiS in space.
As already mentioned in Sec.~\ref{sec:introduction},
SiS has been detected in envelopes around evolved stars and in outflows driven by protostars.

In envelopes around evolved stars, SiS is thought to be formed under thermochemical equilibrium
at the high density and high temperature conditions of the stellar photosphere.
SiS is then injected into the expanding envelope,
where it may vary its abundance due to different types of processes such as shocks induced
by the periodic pulsation of the star, interaction with dust grains, and photoprocesses
driven by the UV radiation emitted by nearby hot stars.
In this sense, the reaction of destruction with O atoms studied here may play
an important role in regulating the SiS abundance. In circumstellar envelopes,
oxygen atoms can be released by the photodissociation of CO, SiO, and H$_2$O. Although
their relative abundances vary depending on the oxigen- or carbon-rich character,
these molecules are the main oxygen reservoirs.
 Photodissociation of these molecules does certainly
takes place in the outer regions of the envelope, which are no longer shielded against UV
photons by the circumstellar dust, but it may also take place in the inner regions
of clumpy envelopes \citep{Agundez2010}. Since the reaction between SiS and O is rapid,
it may provide an efficient way to destroy SiS and recycle the trapped silicon into SiO.
In this line, it is worth noting that interferometric maps of SiO and SiS in the C-star envelope IRC\,+10216
indicates that SiS emission disappears at shorter distances from the star than SiO \citep{Velilla-Prieto2015}.
One possible explanation could be that at moderately short distances from the star,
the amount of O atoms is high enough to efficiently destroy SiS.

In protostellar outflows the situation is different as SiS probably forms from the silicon
released during dust grain disruption. Here we aim to evaluate whether in these environments
the reaction between atomic silicon and SO can be an efficient source of SiS and whether
the reaction between SiS and atomic oxygen can be an important destruction process of SiS which
could potentially explain the paucity of SiS detections in outflows. Protostellar outflows lead
to the generation of shocks between the ejected material and the surrounding quiescent envelope.
During the shock, density and temperature are drastically enhanced and, depending on the shock strength,
disruption of dust grains and dissociation of molecules can happen.
After the passage of the shock, density and temperature relax and molecule formation takes place.
Dedicated models have been constructed to treat this process in detail
(see, e.g., \citealt{Schilke1997,Gusdorf2008}), although they usually focus on SiO and not SiS.

\begin{table}
\caption{Selected reactions and associated rate coefficients} \label{table:reactions}
\centering
\begin{tabular}{lcrrr}
\hline \hline
\multicolumn{1}{l}{Reaction} & \multicolumn{1}{c}{$\alpha$} & \multicolumn{1}{c}{$\beta$} & \multicolumn{1}{c}{$\gamma$} & \multicolumn{1}{c}{Ref.} \\
 & \multicolumn{1}{c}{(cm$^3$ s$^{-1}$)} & & \multicolumn{1}{c}{(K)} & \\
\hline
SiS + O $\rightarrow$ SiO + S & $9.53\times10^{-11}$ & 0.29 & -32 & (1) \\
Si + SO $\rightarrow$ SiO + S & $1.53\times10^{-10}$ & $-0.11$ & 32 & (1) \\
Si + SO $\rightarrow$ SiS + O & $1.77\times10^{-11}$ & 0.16 & -20 & (1) \\
Si + SO$_2$ $\rightarrow$ SiO + SO & $1.00\times10^{-10}$ & 0.00 & 0 & (2) \\
Si + SO$_2$ $\rightarrow$ SiS + O$_2$ & $1.00\times10^{-10}$ & 0.00 & 0 & (2) \\
SiH + S $\rightarrow$ SiS + H & $1.00\times10^{-10}$ & 0.00 & 0 & (3) \\
SiH + S$_2$ $\rightarrow$ SiS + SH & $1.00\times10^{-10}$ & 0.00 & 0 & (3) \\
\hline
\end{tabular}
\tablecomments{Rate coefficient is given by $\alpha$ ($T$/300)$^\beta$ $\exp(-\gamma/T)$.}
\tablerefs{(1) This study. Temperature range of validity is 50-2500 K for SiS + O and 100-2500 K for Si + SO.
  (2) Assumed value for exothermic and spin-allowed reaction.
  (3) Assumed value for barrierless reaction studied by \citet{Rosi-etal:18}.}
\end{table}

Here we have carried out a simple time-dependent chemical model
to explore the influence of the SiS + O and Si + SO reactions on the chemistry of SiS in a post-shocked gas.
These calculations do not aim to model a particular shocked outflow,
but to evaluate how does SiS abundance behaves during molecule formation after the passage of a shock.
For this simple model, we consider that density and temperature remain constant with time.
We adopted a density of H nuclei of 10$^8$ cm$^{-3}$, which makes molecule formation to take place within a few hundreds of years.
The main effect of the density is on the overall chemical timescale, which is shortened in denser gas and lengthened at lower densities.
We consider three different gas kinetic temperatures:
100, 300, and 500 K. We assume that right after the shock passage,
all molecules have been dissociated so that the initial composition of the post-shocked gas consists of neutral atoms with solar elemental abundances.

We use a large chemical network, with $\sim$500 gas-phase species
composed of the elements H, C, O, N, S, Si, and P and linked by $\sim$8000 reactions,
previously used to study the chemistry of IRC\,+10216 \citep{Agundez2017}.
The rate coefficients of most reactions involving SiS as reactant or product in this network
are mere guesses based on Si/C and S/O chemical analogies (e.g., \citealt{Willacy1998}).

We have checked various potentially important SiS-forming reactions involving atomic silicon as reactant.
Excluding endothermic and spin-forbidden reactions,
we identify that Si + SO and Si + SO$_2$ are probably the most efficient ones.
As discussed in the previous section, the reaction between Si and SO has two open channels,
one that yields SiO, which is the most exothermic and rapid,
and another one resulting in SiS, which we also find to be moderately rapid.
In the case of the reaction between Si and SO$_2$,
we assume that the channels leading to SiO and SiS are equally rapid,
with a rate coefficient of 10$^{-10}$ cm$^3$ s$^{-1}$.
This same value is adopted for the SiS-forming reactions SiH + S and SiH + S$_2$,
which have been calculated to be barrierless by \citet{Rosi-etal:18},
although in our calculations these reactions are not important routes to SiS.
{ The reason is that SiH is not
  efficiently formed in gas phase from Si because the reaction between Si and H$_2$ to yield SiH
  is endothermic by about 35 kJ/mol, and thus too low even at 500K}
The reaction of SiS with atomic oxygen, which is calculated to be rapid, is also included.
The rate coefficients adopted for these reactions are summarized in Table~\ref{table:reactions}.

\begin{figure}
\centering
\includegraphics[angle=0,width=\columnwidth]{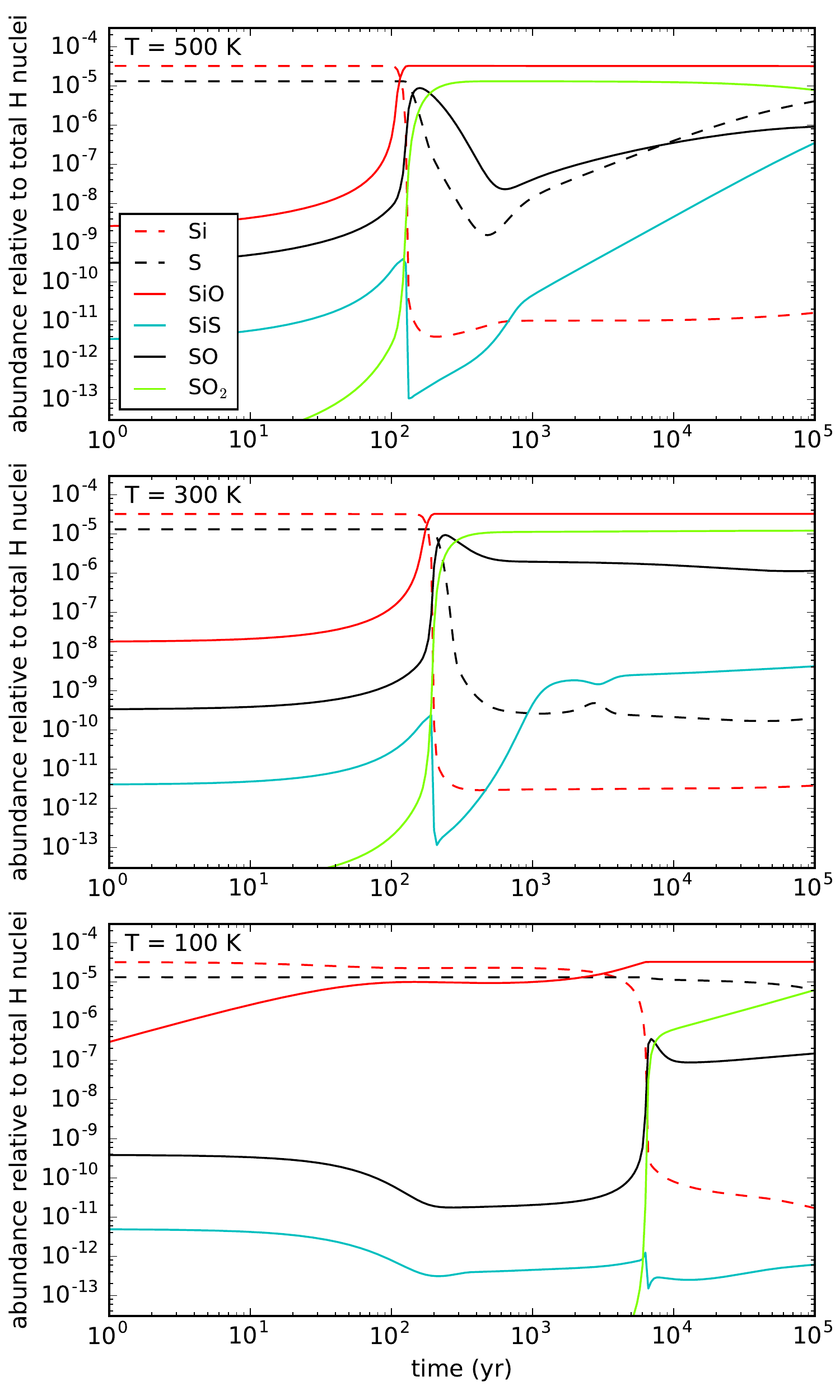}
\caption{Calculated abundances of SiO, SiS, and related species as a function of time for postshock conditions.
  The assumed density is 10$^8$ cm$^{-3}$ and adopted temperatures are 100 K (lower panel),
  300 K (middle panel), and 500 K (upper panel).} \label{fig:postshock}
\end{figure}

In Fig.~\ref{fig:postshock} we show the time evolution of the abundances of SiS, SiO,
and some other interesting species. We first note that SiO is formed rapidly from atomic Si reacting
with O$_2$ and OH at any temperature. This is also found in previous models of shocked protostellar
outflows (e.g., \citealt{Schilke1997,Gusdorf2008}).
The formation of SiS, on the other hand, is found to be substantially less favored than that of SiO,
both regarding the maximum abundance reached and the velocity of formation.
More concretely, SiS is formed efficiently only at temperatures higher than 100 K
and its formation is not as rapid as that of SiO.
At temperatures of a few hundreds of Kelvin,
the formation of SiS at early postshock times occurs at the expense of atomic silicon,
in a similar way as occurs for SiO.
The abundance of SiS however remains 2-3 orders of magnitude below that of SiO because,
on the one hand, the sulfur precursors of SiS (SO and SO$_2$)
are not as abundant as the oxygen precursors of SiO (O$_2$ and OH), and on the other,
because SiS is efficiently destroyed by atomic oxygen forming SiO.
At late postshock times, SiS can become quite abundant,
where the main formation reactions are Si + SO and Si + SO$_2$.
We note that given the simplicity of the calculations presented here,
it is not advisable to extract quantitative conclusions on abundances and time scales.
Nevertheless, we can extract useful information from these calculations:
(1) SiS forms with a lower abundance and later than SiO,
(2) SiS is efficiently destroyed through reaction with O, and
(3), the main SiS-forming reactions are Si + SO and Si + SO$_2$.

\section{Conclusions} \label{sec:conclusions}

We have calculated the {\it ab initio} PESs of the SiOS system to compute the rate coefficients of the reaction
of destruction of SiS with atomic O and of the SiS-forming reaction Si + SO in the temperature range 50-2500 K.
We find that the reactions SiS + O and Si + SO are rapid,
with rate coefficients of a few times 10$^{-10}$ cm$^3$ s$^{-1}$,
almost independent of temperature.
In the reaction between Si and SO, production of SiO is favored over formation of SiS,
with a 5-7 times larger rate coefficient. The reaction of SiS with O atoms can
play an important role in destroying SiS in envelopes around evolved stars and definitively
is an efficient destruction pathway for SiS in protostellar outflows.
Through a simple chemical model we find that in this latter type of environments
SiS is formed with a lower abundance than SiO and that the reactions of formation are
Si + SO and Si + SO$_2$. Dedicated observational searches for SiS and detailed shock
models including SiS chemistry are needed to better understand the differential behavior
of SiO and SiS in interstellar and circumstellar clouds.

\acknowledgments

The research leading to these results has received funding from
the European Research Council (ERC Grant 610256: NANOCOSMOS)
and from Spanish MINECO through grants FIS2014-52172-C2, FIS2017-83473-C2,
and AYA2016-75066-C2-1-P.
M.A. also acknowledges funding support from the Ram\'on y Cajal programme
of Spanish MINECO (RyC-2014-16277).


 \end{document}